# Substrate-Independent Growth of Atomically Precise Chiral Graphene Nanoribbons

*Dimas G. de Oteyza,*[1,2,3,*]*Aran Garcia-Lekue,*[1,2] *Manuel Vilas-Varela,*[4] *Nestor Merino,*[1,5] *Eduard Carbonell-Sanromà,*[5] *Martina Corso,*[2,3,5] *Guillaume Vasseur,*[1,3] *Celia Rogero,*[1,3] *Enrique Guitian,*[4] *Jose Ignacio Pascual,*[2,5] *J. Enrique Ortega,*[1,3,6] *Yutaka Wakayama,*[7] *Diego Peña* [4,*]

[1] Donostia International Physics Center (DIPC), Paseo Manuel Lardizabal 4, 20018 San Sebastián, Spain

[2] Ikerbasque, Basque Foundation for Science, 48011 Bilbao, Spain

[3] Centro de Física de Materiales (CSIC/UPV-EHU) -Materials Physics Center, Paseo Manuel Lardizabal 5, 20018 San Sebastián, Spain

[4] Centro de Investigación en Química Biolóxica e Materiais Moleculares (CIQUS) and Departamento de Química Orgánica, Universidade de Santiago de Compostela, 15782, Spain

[5] CIC nanoGUNE, Avenida de Tolosa 76, 20018 San Sebastián, Spain

[6] Departamento de Fisica Aplicada I, Universidad del Pais Vasco, 20018 San Sebastián, Spain

[7] International Center of Materials Nanoarchitectonics, National Institute for Materials Science,

1-1 Namiki, Tsukuba 305-0044, Japan

tunneling microscopy, core-level photoemission, density functional theory.

ABSTRACT. Contributing to the need of new graphene nanoribbon (GNR) structures that can be synthesized with atomic precision, we have designed a reactant that renders chiral (3,1)-GNRs after a multi-step reaction including Ullmann coupling and cyclodehydrogenation. The nanoribbon synthesis has been successfully proved on different coinage metals, and the formation process, together with the fingerprints associated to each reaction step, has been studied combining scanning tunnelling microscopy, core-level spectroscopy and density functional calculations. In addition to the GNR´s chiral edge structure, the substantial GNR lengths achieved and the low processing temperature required to complete the reaction grant this reactant extremely interesting properties for potential applications.

Graphene nanoribbons (GNRs) are drawing enormous interest, partly due to their attractive electronic properties.[1,2] Those properties vary dramatically with changes in the nanoribbon's atomic structure in terms of width,[3-5] crystallographic symmetry,[6,7] dopant heteroatoms[8-13] and edge termination.[14] Moreover, the electronic properties can be modulated even further by the appropriate design of GNR heterostructures.[12,15-17] This enormous tunability of electronic properties is thus extremely promising for next-generation nanoelectronic and optoelectronic devices.[18,19] However, the high susceptibility of those properties to minimum changes in the GNR structure also remarks the stringent need for atomic precision in GNR synthesis. With the advent of bottom-up synthesis,[1,2,20] increasingly high hopes are being placed on this approach, but the field is still at its birth. Thus, although a large pool of GNRs with different edge orientations, widths or heteroatoms (and heterostructures thereof) should become available to really allow for the envisioned breakthroughs in nanoelectronics and the development of full GNR-based circuitry, so far only few GNRs have been successfully synthesized with the required selectivity and precision.[1,2,8-10,20-24]

To date, the most widely studied nanoribbon is the armchair-oriented GNR with 7 dimer lines across its width (7-AGNR) that grows from 10,10'-dibromo-9,9'-bianthracene (reactant **1** in Figure 1a) in a multistep reaction including dehalogenation, polymerization (also known as Ullmann coupling) and cyclodehydrogenation.[20,25,26] The synthesis of 7-AGNR has been shown to work reproducibly on substrates like Au(111),[20] Au(110)[27] or Ag(111).[20,28] Surprisingly, the same reactant **1** designed to render AGNRs turned out to form chiral (3,1)-GNRs on Cu(111) (Figure 1).[29,30] This result has been subject to debate,[31-34] since the polymerization does not involve the carbon atoms attached to bromines. However, the debate has been recently settled by unambiguous high-resolution imaging of the resulting bonding structure.[35] These results mirror a very system-specific reaction mechanism not translatable to other substrates, based on the surface-catalyzed, selective activation of particular C-H bonds. In fact, similar results were obtained from **1** and its non-halogenated sister molecule, the latter producing slightly longer GNRs.[35] This implies an absent or even negative impact of the halogenation of **1** for GNR growth on Cu(111).

Nevertheless, those results have inspired this work, in which we introduce monomer **2**, specifically designed to obtain (3,1)-GNRs in a more efficient manner (Figure 1). In particular, the bromine atoms in positions 10,10' of reactant **1** have been shifted in reactant **2** to positions 2,2', which are the positions selectively activated on Cu(111) and through which the polymerization preceding the (3,1)-GNR formation takes place. The same structure can thus be obtained from **2** following a conventional Ullmann coupling/cyclodehydrogenation sequence. This has been shown to work independently of the used substrate, as proved with growth studies on Au(111) and Ag(111). A similar study on Cu(111) is out of the scope of this paper due to the readily proved growth of (3,1)-GNRs from **1** (and even from non-halogenated precursors) on that substrate. However, also on Cu(111) we show how changing the halogen functionalization site, and thereby changing the polymerization mechanism from a selective C-H bond activation to Ullmann coupling, is still a significant advancement by greatly increasing the resultant GNR length.

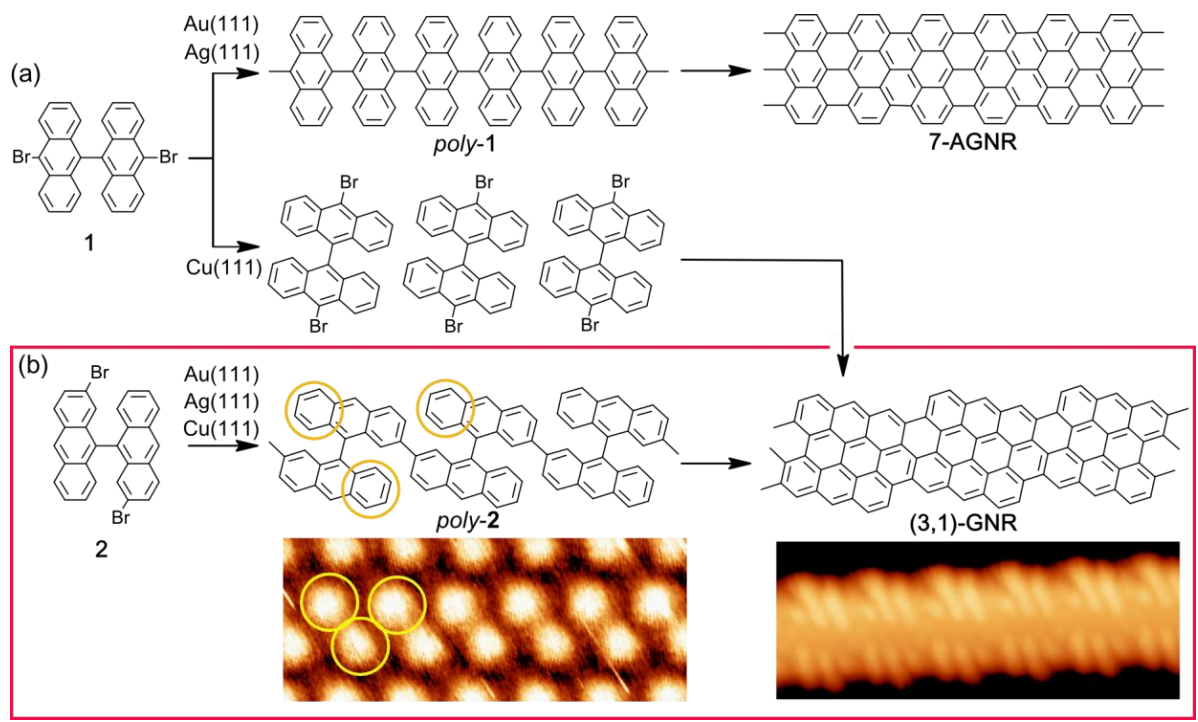

**Figure 1.** (a) Schemes of the chemical reactions of precursor **1** on various metallic surfaces. On Au(111) and Ag(111) it affords armchair GNRs. On Cu(111), different groups report formation of 7-AGNRs or chiral (3,1)-GNRs. (b) Our work (highlighted with the red line) reports the transformation of reactant **2** into chiral GNRs independently of the substrate [Au(111), Ag(111 and Cu(111)]. Associated STM images are shown for *poly*-**2** after initial polymerization by Ullmann coupling (5.6 nm × 2.2 nm, I=0.09 nA, U=1.5 V), as well as for the final (3,1)-GNR after cyclodehydrogenation (5.6 nm × 2.2 nm, I=0.2 nA, U = -650 mV), both on Au(111). Steric hindrance causes *poly*-**2** to be non-planar. The high parts (circled in yellow) are correspondingly marked in the polymer's wireframe structure above.

## Results and discussion

Key to this work is the synthesis of bianthracene reactants with adequately chosen bromine atom positions. We prepare 2,2'-dibromo-9,9'-bianthracene (**2**) starting from phthalic anhydride (**3**), following the four-steps synthetic route shown in Figure 2 (see supporting information, Scheme

S1, Figures S1 and S2 for details). The key reaction in this protocol is the reductive coupling of bromoanthrone **4** promoted by Zn.[36]

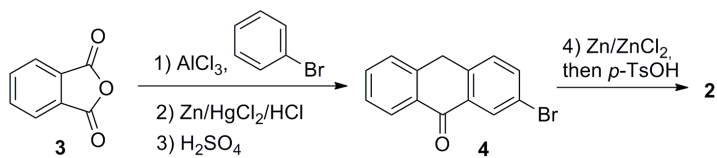

**Figure 2**. Synthetic route to obtain dibromobianthracene **2** from phthalic anhydride (**3**).

Starting from reactant **2**, the reaction pathway associated with the surface-supported synthesis of (3,1)-GNRs is closely related to the well-known transformation of **1** into 7-AGNRs. That is, in a first step the molecules polymerize by Ullmann coupling into *poly*-**2**. This polymer is a highly non-planar molecular structure in which the steric hindrance between the H-atoms in neighbouring anthracene units drives their alternating tilting (Figure 1b). As a consequence, the polymer's imaging by scanning tunnelling microscopy (STM) displays a sequence of protruding features that we associate with the up-pointing ends of the anthracene units. This correspondence is highlighted by yellow circles in the wireframe chemical structure and in the STM image. In a following reaction step, cyclodehydrogenation sets in and *poly*-**2** transforms into the planar (3,1)-GNR structure, as can be directly discerned in the high resolution STM images in Figure 1b and Figure S3.

Figure 3 summarizes, as observed by STM, the growth process of (3,1)-GNRs on top of Au(111) and Ag(111). On either substrate, the images correspond to the same sample at different stages of its growth: after deposition of **2** on substrates held at RT, after annealing to 150 °C, and after annealing to 205 °C.

We first focus on Au(111). After RT deposition, the molecules aggregate into islands of linear structures formed by a zig-zag arrangement of protrusions comparable to those expected from *poly*-**2** (Figures 3a and 3b). Upon annealing to 150 °C we observe clear changes in the sample's topology. It is difficult to discern changes in the STM contrast within the linear structures (see supplementary information for details). However, their overall length substantially increases and the spacing between them (perpendicular to the structure's long axis) becomes less regular and decreases the minimum distance (Figures 3c and 3d). Annealing to 205 °C brings about more notorious changes, displaying arrays of planar structures clearly recognizable from the edge topology as (3,1)-GNRs (Figures 3e and 3f).

In the case of Ag(111), two distinct sections are observed after molecular deposition at RT. On the one hand we find regions of ordered, linear structures packed side by side (Figure 3g). The linear structures are imaged again as zig-zagging protrusions (Figure 3h). Instead, other regions display a disordered arrangement of adsorbates with increased mobility and a much larger apparent height (~2.7 Å vs. ~1.8 Å). The areal ratio between these two different sections is approximately one to one (Figure 3h). Annealing to 150 °C brings about the growth of the ordered, linear structures at the expense of the disappearing disordered regions (Figures 3i and 3j). As opposed to the findings on Au(111), on Ag(111) the arrangement within the ordered arrays of linear structures remains unchanged after this annealing. Annealing the sample to 205 °C triggers the cyclodehydrogenation and thereby the ultimate formation of (3,1)-GNRs (Figures 3k and 3l).

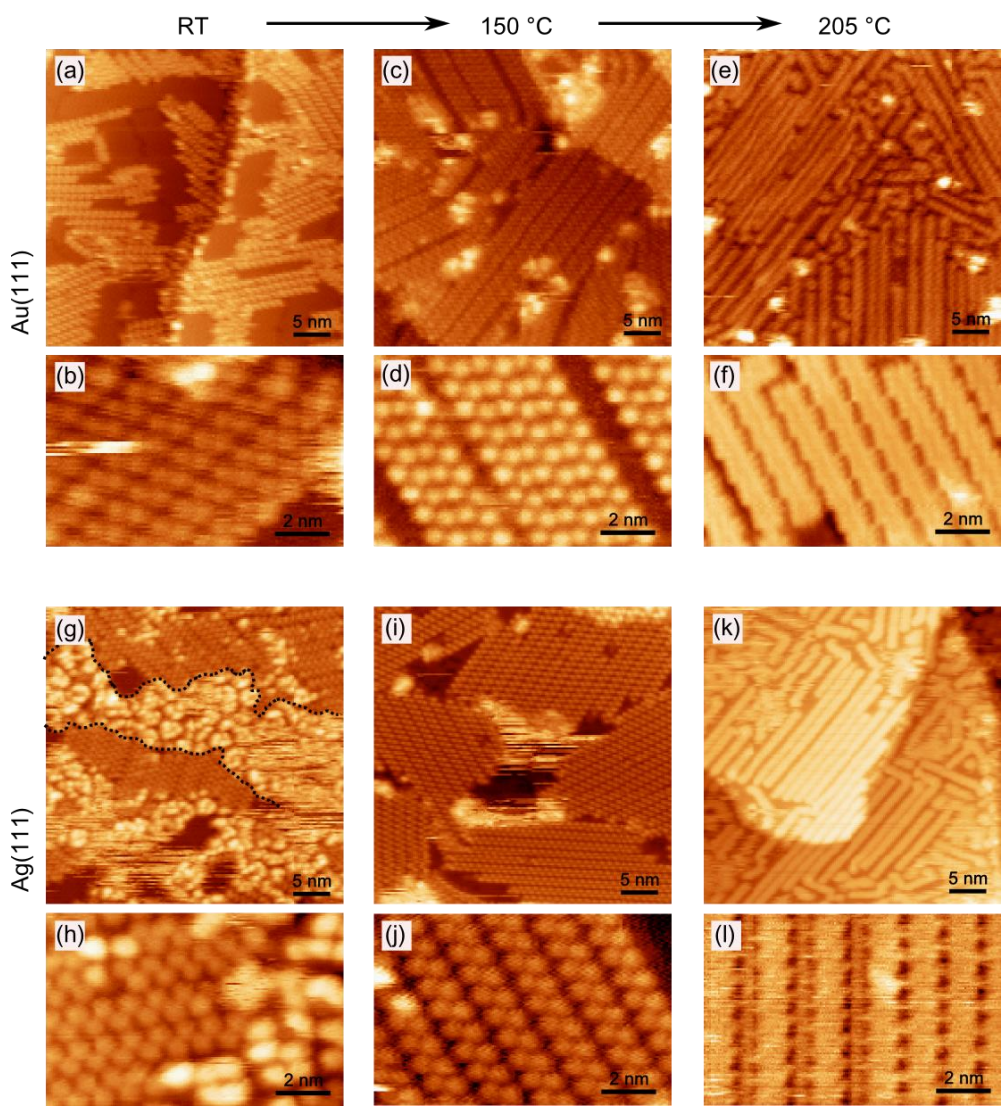

**Figure 3.** Large scale (36 nm × 36 nm) and smaller scale (10 nm × 6.5 nm) STM images on Au(111) (a-f) and Ag(111) (g-l) of the same samples at different growth stages: after deposition on substrates held at room temperature (a,b,g,h), after annealing to 150 °C (c,d,i,j) and after annealing to 205 °C (e,f,k,l),. STM imaging parameters are: (a) I=0.086 nA, U=1.4 V, (b) I=0.086 nA, U=1.5 V, (c) I=0.16 nA, U=1.76 V, (d) I=0.16 nA, U=1.76 V, (e) I=0.16 nA, U=0.47 V, (f) I=1.29 nA, U=-0.13 V, (g) I=0.09 nA, U=1.5 V, (h) I=0.06 nA, U=-2.02 V, (i) I=0.36 nA, U=1.76 V, (j) I=0.36 nA, U=-1. 6 V, (k) I=0.42 nA, U=1.07 V, (l) I=0.19 nA, U=-0.13 V.

Complementary information on the chemical transformation process is obtained from core level photoemission spectroscopy (XPS) measurements. As in the STM experiment, molecules were deposited on Au(111) and Ag(111) substrates held at room temperature. The samples were then annealed stepwise while monitoring their Br 3p and C 1s core level spectra. The data are summarized in Figure 4.

On Au(111), the molecules remain intact upon deposition at RT and only start showing chemical changes for substrate temperatures of around 125 °C. As the temperature increases above that threshold, the most evident change in the core level spectra is a pronounced shift of the Br 3p peaks to ~2 eV lower binding energies. This effect is well known from other studies on Ullmann coupling of different precursors and relates to the dehalogenation process and the new chemical environment as Br detaches from the organic molecule and binds to the metallic surface.[27,37,38] Concomitant, the C 1s peak displays a smaller shift (~0.3 eV) to lower binding energies. Similar C 1s shifts have also been observed in previous studies on Ullmann coupling with different precursors and surfaces, for which a variety of explanations have been given: (i) bond formation between the C atoms hosting the generated radicals and the substrate atoms or adatoms,[27,37] (ii) a change in the supramolecular assembly prior to dehalogenation that brings about changes in the interaction with the substrate,[38] or (iii) a change of work function caused by the chemisorption of Br to the substrate.[37] We discard the first because on Au the formation of organometallic compounds is disfavored and the molecules are known to polymerize as the radicals are formed.[39,40] We also discard the second because we observe the C 1s and Br 3p shifts simultaneously in a correlated way. Thus, we ascribe the observed C 1s shift to the change of work function generated by the metal-bound Br atoms. Besides, this is further supported by the changes observed in the core levels as the temperature is increased further: as Br desorption starts to set in, the C 1s level shifts in the opposite direction, towards higher binding energy. A similar shift on closely related systems has also been ascribed to different reaction processes like the transformation from an organometallic to a polymer phase[27,41] or cyclodehydrogenation.[37] In addition to the chemical change, these reactions also cause an alteration in the molecule-substrate distance that may in turn additionally affect the core-hole screening effects and thereby the core level spectra. However, we can again discard these justifications for our system because we know the polymerization and cyclodehydrogenation to occur at lower temperatures. Altogether, XPS on Au(111) thus shows that the structures observed on Au(111) at RT are a non-covalent self-assembled supramolecular arrangement, polymerizing only upon annealing above 125 °C, in line with the overall sample topology changes observed by STM at 150 °C and described above.

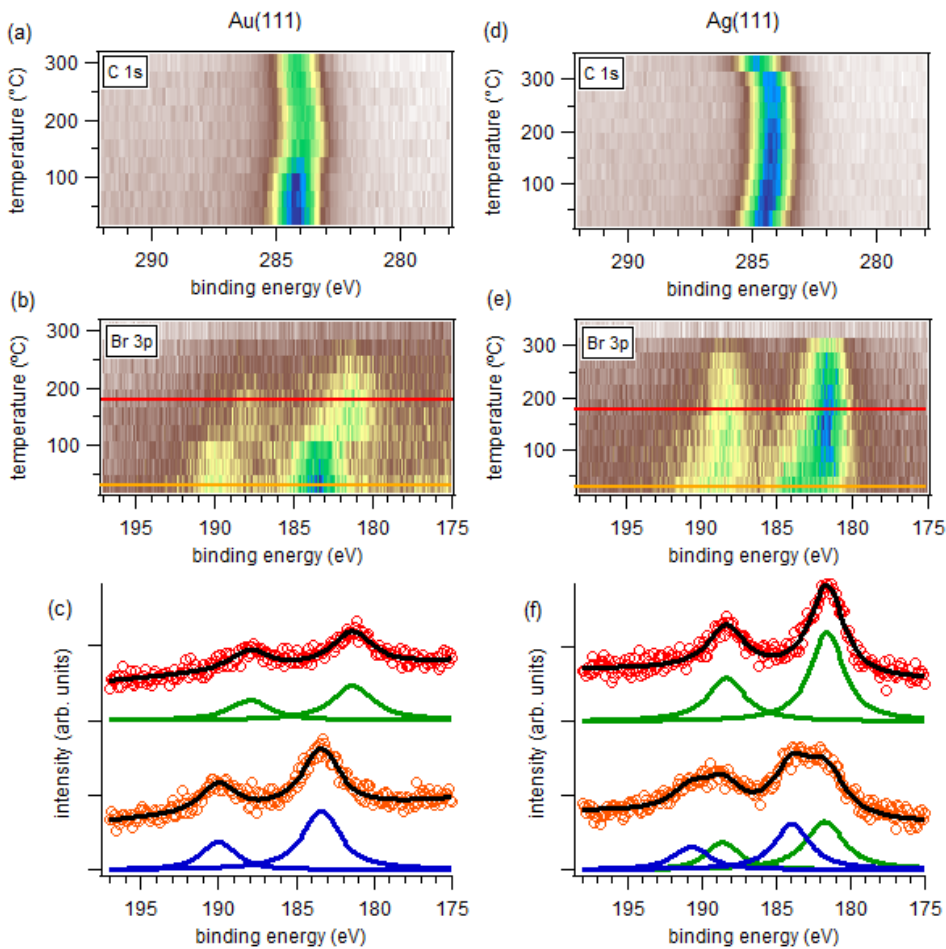

**Figure 4.** Photoemission spectra of the C 1s core levels of **2** deposited on (a) Au(111) and (d) Ag(111) held at room temperature and their evolution as a function of sample annealing temperature. Similar measurements of the Br 3p core levels are shown in panels (b) and (e). Panels (c) and (f) depict Br 3p spectra, together with their associated fits

(blue and green lines correspond to organic and metal-bound Br components, respectively), of two representative temperatures marked with the colored lines in (b) and (e), respectively. The spectra are shifted along the intensity axis for better comparison.

On Ag(111) the molecules are readily partially dehalogenated upon deposition at RT. This is clearly observed in the Br 3p core level spectra, which evidence the coexistence of organic and metal-linked Br atoms in a 1:1 ratio. As the temperature is increased, the metal-bound Br increases at the expense of the organic Br. From the correlation with the STM observations we can readily ascribe the pristine precursors to the disordered structures found at RT, and the dehalogenated molecules to be the building blocks forming the ordered structures. As in Au(111), the dehalogenation brings about a minor shift to lower binding energy in the C 1s spectrum, which shifts back again as the Br is desorbed at higher temperatures. Again, we ascribe the dominating effect behind the C 1s shifts to the changes in work function as Br binds or leaves the metal surface. However, spectroscopy-wise this leaves the question open as to what is the nature of the linear, ordered structures formed by the dehalogenated precursors. Do those radical species link covalently or *via* metal-organic coordination, as is commonly the case on Ag at temperatures below ~150 °C? In the absence of clear spectroscopic fingerprints, the answer will be given based on periodicity analysis along the one-dimensional structures, organometallic structures typically having significantly larger periodicities than polymers.[40,42,43]

DFT calculations for free standing structures predict periodicities of 9.52 Å for the metal-organic chain (Figure 5c), 8.21 Å for the polymer (Figure 5b) and of 8.96 Å for the flat (3,1)-GNR (Figure 5a), the latter in good agreement with the value of 8.89 Å that results from assuming an undisturbed graphene lattice for the GNR. Our STM measurements reveal the periodicity of the GNRs to be 9.0±0.8 Å, excellently fitting the calculations. For the non-planar structures we observe the same periodicity at RT and after annealing to 150 °C, with an average value of 8.2±0.7 Å. The value is similar to that of *poly*-**2** on Au(111) (8.3±0.6 Å) and in excellent agreement with a polymeric phase, the error margins remaining clearly under the metalorganic periodicity (Fig. 5d). Thus, in spite of the flexibility associated with the non-planarity of *poly*-**2** and the metalorganic chain, which may introduce minor uncertainties in the calculated periodicities, the results still allow us to conclude that the polymer is directly formed after dehalogenation.

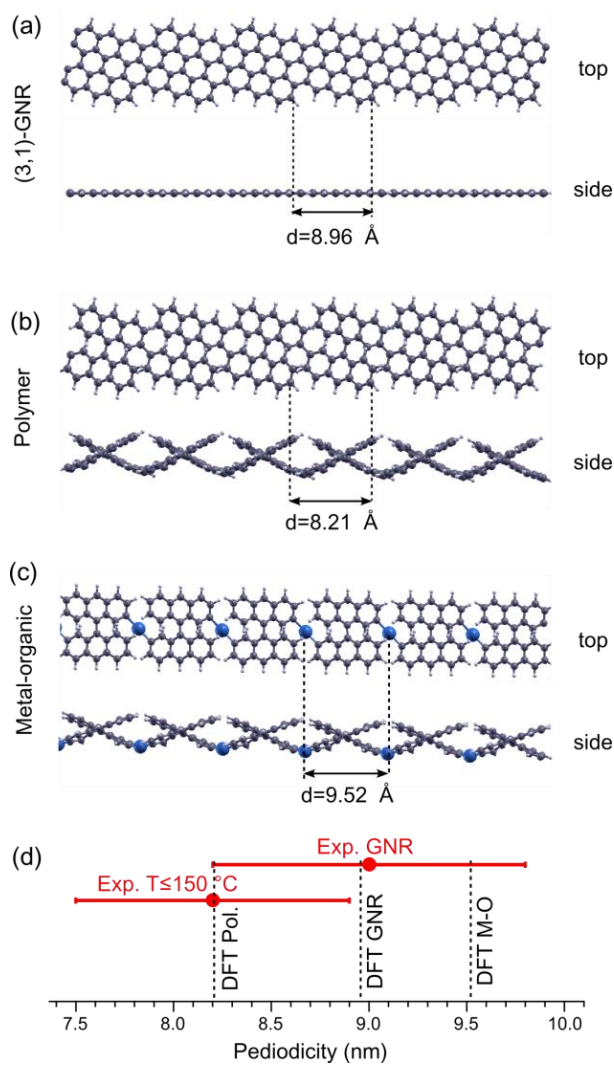

**Figure 5.** Relaxed structures for (a) free-standing (3,1)-GNRs, (b) *poly*-**2** and (c) metal-organic (M-O) chains. (d) Comparison of the periodicities of the calculated structures with those measured experimentally for the GNRs and for the non-planar structure on Ag(111). The latter fits the polymer period and its error margin (shown as the standard deviation) is well below the periodicity of the metal-organic chain.

As only reported with few other systems,[44] we thus observe a covalent polymer formation readily at room temperature, instead of the metal-organic intermediates reported with most precursors linking through Ullmann coupling on Ag(111) and Cu(111).[37,39-42] The reason behind this may be sought in the different coordination geometry imposed by the non-planar precursors. As observed in most previous Ullmann coupling studies on Ag surfaces, Ag atoms present linear coordination geometry,[39,40,42] while the anthracene units hosting the generated radicals upon dehalogenation of **2** display a strong tilt with respect to the substrate plane due to the steric congestion within the organic backbone (Fig. 5c). Under these circumstances, and based on our experimental observations, we conclude that the metal-organic intermediate is not sufficiently stable and the reaction directly proceeds towards the polymeric phase.

As noted above, at 205 °C the GNRs are fully cyclodehydrogenated both on Ag(111) and Au(111). Similarly low cyclodehydrogenation temperatures (177 °C) are reported for reactant **1** on Ag(111),[28] but higher temperatures, similar to those required on Au(111), are needed to really form GNRs (377 °C).[20,28] On Cu(111), the required temperature for fully cyclodehydrogenated GNRs to be formed is 250 °C.[33,35] Most remarkably, reactant **2** renders fully dehydrogenated GNRs at significantly lower temperatures than **1** even on the low reactivity Au(111) surface. This surprisingly large change relates to the substantially altered strain in the two polymer structures, since sterically induced strain is known to weaken the involved C-H bonds and thereby lower the cyclodehydrogenation barriers.[45,46] In *poly*-**1**, the anthracene units are linked covalently along their short axis by a bond that allows free rotational movement with respect to their neighbors. This freedom results in alternatively tilted anthracene units along the polymer backbone so as to minimize the steric hindrance from opposing H atoms. Instead, the anthracene units within *poly*-**2** are linked covalently to their neighbors both along their long and short axes. Thus, although the anthracene units still display the same alternative tilt to reduce the steric hindrance, the covalent bonds along the long anthracene's axes limit the structure's rotational freedom, resulting in a substantially strained geometry. It is this strain opposing the anthracene's tilting which favors the planarization of the structure and thus reduces the cyclodehydrogenation temperature threshold regardless of the substrate.

Both on Ag(111) and Au(111), the cyclodehydrogenation threshold is between 150 °C, at which no cyclodehydrogenation is observed, and 205 °C, at which the whole sample has readily become fully planar (Figure 3). On Ag(111), where polymerization readily starts at room temperature, there is still a substantial temperature gap before the onset of cyclodehydrogenation. However, on Au(111) the threshold temperatures for polymerization (~125 °C) and cyclodehydrogenation are in close proximity. This may have an impact on the growth process and the resulting GNRs, since liberated H in the cyclodehydrogenation process could quench the available radicals and terminate the polymerization. To shed light on this issue we have increased the number of sampling temperatures, figuring the cyclodehydrogenation onset on Au(111) to be below 175 °C, at which most of the sample has readily become a planar GNR but some of the polymer units still remain unreacted (inset in Figure 6). A systematic GNR length analysis of samples as a function of the substrate temperature upon reactant deposition is shown in Figure 6, all samples having coverages of around 0.8 ML. For temperatures below the cyclodehydrogenation threshold, a second annealing step to 205 °C was applied to the sample for GNR formation before performing the length analysis.

Representative distributions for selected temperatures are shown in Figure 6a, making immediately obvious that high temperatures narrow the distribution significantly and prevent formation of long GNRs. Because of the asymmetric length distribution we take the median length as a representative value and plot it *versus* substrate temperature upon first deposition (Figure 6b). We observe an important drop in the length with increasing temperature once the cyclodehydrogenation threshold is passed. Under this scenario, radical step growth and cyclodehydrogenation take place simultaneously. Thus, radical quenching by liberated H atoms competes with the radical step-growth polymerization. Deposition on Au held at room temperature and subsequent annealing to cyclo-dehydrogenation temperatures suffers from the same effect, since the precursors on the surface remain intact at room temperature and both polymerization and cyclodehydrogenation occur during the same subsequent annealing process. However, the length-limiting effect is less pronounced, among other reasons due to the finite heating rate. Longest GNRs are obtained at substrate temperatures that first activate polymerization, only to form the GNRs in a subsequent annealing process. Under these circumstances, GNRs in excess of 30 nm can be easily obtained, well beyond the longest (3,1)-GNRs obtained from **1** on Cu(111).[30] Moreover, additional studies to maximize GNR lengths by optimizing surface coverages or heating rates may bring about even further improvements in the future.

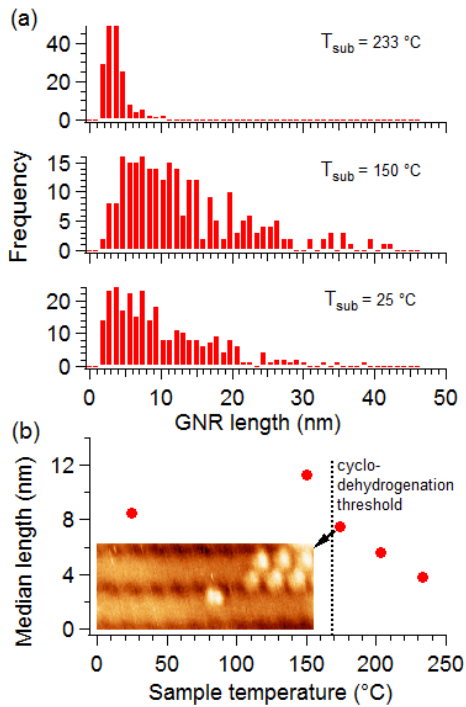

**Figure 6.** a) Length distribution of GNRs grown on Au, at coverages close to the full monolayer, for various different substrate temperatures upon first deposition. For temperatures below the cyclodehydrogenation threshold (T < 174 °C), a second annealing step to 205 °C was applied to the sample for GNR formation by cyclodehydrogenation. b) Median length for each substrate temperature. The inset depicts an STM image (7.4 nm × 2.9 nm, I = 0.16 nA, U = 0.47 V) of a sample deposited at 174 °C revealing a mostly, but not yet fully, cyclodehydrogenated structure. This value has been thus taken as the cyclodehydrogenation threshold temperature.

Lastly, we have confirmed the suitability of this molecule to form longer (3,1)-GNRs than **1** also on Cu(111). This is shown in Figure 7 and underlines the great advancement provided by this new GNR precursor. In addition to the (3,1)-GNR formation, a concomitant etching of triangular holes into the remaining uncovered Cu(111) surface is observed, lined along their sides by Br atoms (Figure 7). A detailed study and description of this process, however, is beyond the scope of this paper. We also want to remind that the GNR length analysis of this sample on Cu(111) (Figure 7c) should not be compared with that on Au(111) (Figure 6), since the growth was performed in a different chamber with different coverage and a different heating rate, two parameters that may play an important role in the length distribution. But most importantly we want to remark that, different from what occurs with precursor **1** on Cu(111), the precursor **2** allows on the one hand to grow chiral GNRs on different materials not relying on the specific and strong molecule-substrate interaction. On the other hand, on all surfaces studied it forms long GNRs that easily exceed several tens of nanometers, a great advantage for their implementation in actual device structures.

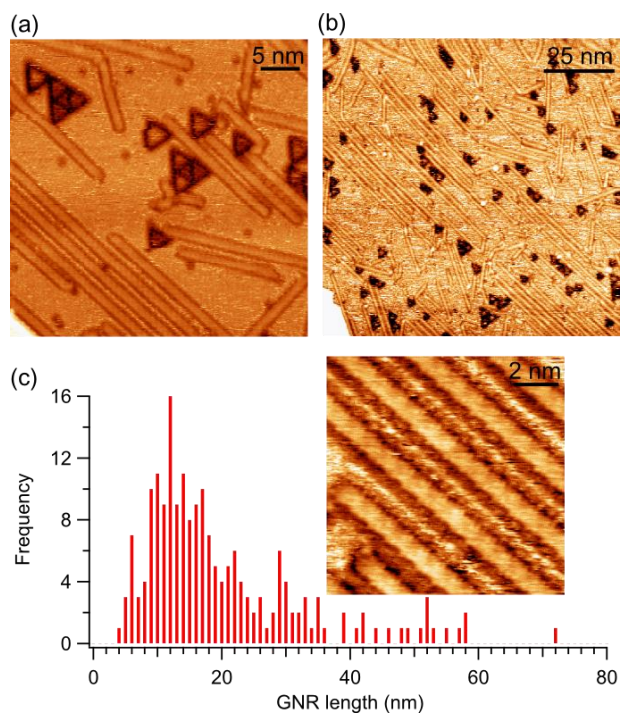

**Figure 7.** (a) $36 \times 36$ nm$^2$ (I = 0.58 nA, U = -400 mV) and (b) $100 \times 100$ nm$^2$ (I = 0.05 nA, U = -200 mV) constant current STM images of (3-1)-GNRs on Cu(111). (c) GNR length distribution as obtained from large-scale images as in (b), showing a significant portion of GNRs to be longer than 40-50 nm and the average and median length values well-above those of GNRs grown from **1** on Cu(111). The inset depicts a $10 \times 10$ nm$^2$ close-up view of the GNRs (I = 1 nA, U = -150 mV).

## Conclusions

Inspired by the previously reported system-specific growth of (3,1)-GNRs on Cu(111) from precursor **1**, we have designed an alternative building block **2** that renders the same (3,1)-GNRs, but now independently of the substrate material. This has been proved on Au(111), Ag(111) and Cu(111), revealing additional advantages of the use of this monomer in the growth of selective and atomically precise GNRs, as is the substantially increased length of the resultant GNRs and the low processing temperature required for their formation. Furthermore, the growth process has been followed in detail combining core-level spectroscopy, scanning tunnelling microscopy and density functional theory calculations, providing a clear correlation between the spectroscopic fingerprints and the different reaction processes, as well as revealing the unusual absence of a meta-stable metal-organic intermediate preceding the covalent polymerization in the Ullmann coupling process on Ag(111).

**Methods**

The various metal (111) surfaces were all prepared by standard sputtering-annealing cycles. Subsequently, the samples were prepared by thermal evaporation of **2** at ~140 °C onto the substrate. Substrate temperature was controlled by resistive heating, and the calibration for the STM experiments was performed by direct measurement of the substrate surface temperature, as a function of the resistive heating current, with a thermocouple fixed and glued to the surface with silver-paint after completion of the experiments (see supplementary information for more details). STM was measured in a commercial UHV system at room temperature, except for the GNR image in Figure 1b, measured in a commercial UHV system at 5 K. WSxM software was used to process all STM images.[47] XPS measurements were performed using a non-monochromatized source. The XPS data were collected by means of a SPECS Phoibos 100 hemispherical electron analyzer, making use of Al K$_\alpha$ x-ray emission.

*Ab-initio* calculations were carried out on free-standing structures using density functional theory (DFT), as implemented in the SIESTA code.[48-50] The optB88-vdW functional, which accounts for non-local corrections, was adopted for the exchange and correlation potential. We employed a double-$\zeta$ plus polarization (DZP) basis set, and a mesh-cutoff of 300 Ry for the real-space integrations. A variable-cell relaxation of the periodic systems was performed until residual forces on all atoms were less than 0.01 eV/Å, and a Monkhorst-Pack mesh with 101x1x1 k-point sampling of the three-dimensional Brillouin zone was used.

Details on the four-steps synthetic route to obtain dibromobianthracene **2** from phthalic anhydride (**3**) are given in the Supporting Information.

ASSOCIATED CONTENT

**Supporting Information**. Details on the reactant synthesis, intramolecular resolution STM imaging, description of the temperature measurement method and its calibration, details on the temperature-dependent STM contrast along linear structures. This material is available free of charge *via* the Internet at http://pubs.acs.org.

AUTHOR INFORMATION

**Corresponding Author**

d_g_oteyza@ehu.es, diego.pena@usc.es

**Author Contributions**

The manuscript was written through contributions of all authors. All authors have given approval to the final version of the manuscript.

## ACKNOWLEDGMENT

The project leading to this publication has received funding from the European Research Council (ERC) under the European Union's Horizon 2020 research and innovation program (grant agreement No 635919), from ICT-FET EU integrated project PAMS (agreement No. 610446), from the Spanish Ministry of Science and Competitiveness (MINECO, Grant Nos. MAT2013-46593-C6-01, MAT2013-46593-C6-2-P, MAT2013-46593-C6-4-P, MAT2013-46593-C6-6-P), FEDER, from the Basque Government (Grant Nos. IT-621-13, IT-756-13 & PI2015-042), from the World Premier International Center (WPI) for Materials Nanoarchitectonics (MANA) of the National Institute for Materials Science (NIMS), Tsukuba, Japan and from JSPS Bilateral Open Partnership Joint Research Project.

**Table of contents**

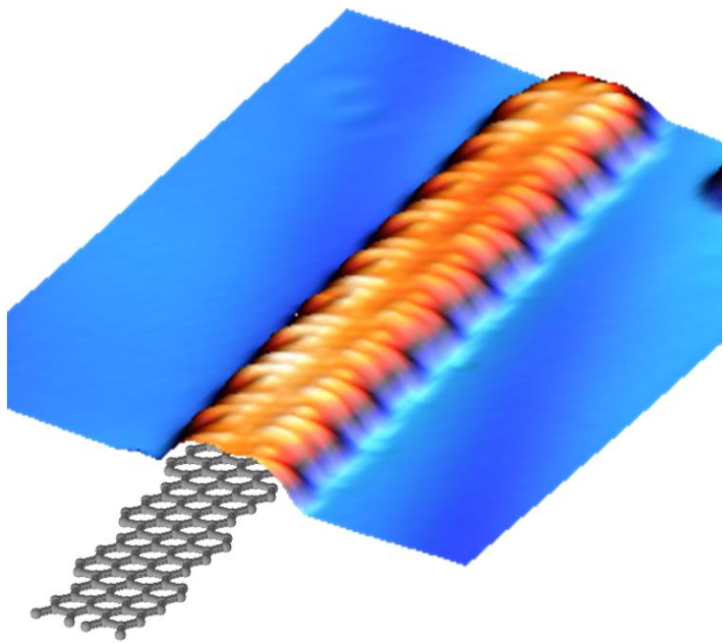